\documentclass[12pt]{article}
\usepackage{amssymb,amsmath}
\usepackage[noblocks]{authblk}
\usepackage[top=0.75in, bottom=0.75in, left=0.75in, right=0.75in, dvips]{geometry}
\usepackage{caption}  
\date{}   

\begin{document}
\textwidth 10.0in       
\textheight 9.0in 
\topmargin -0.60in
\title{Multiple Gauge Fixing Conditions}
\author[]{D.G.C. McKeon\thanks{Email: dgmckeo2@uwo.ca}}
\affil[] {Department of Applied Mathematics, The
University of Western Ontario, London, ON N6A 5B7, Canada}
\affil[] {Department of Mathematics and
Computer Science, Algoma University, Sault St.Marie, ON P6A
2G4, Canada}
\maketitle

\maketitle
\noindent
PACS No.:  11.15 Bt\\
Key Words: gauge fixing, ghosts

\begin{abstract}
We consider how more than one gauge fixing condition can be accommodated within the Feynman path integral both by extending the Faddeev-Popov procedure and the BV approach.  The first order Einstein-Hilbert action in $1 + 1$ dimensions and the massless spin-$\frac{3}{2}$ action are considered.
\end{abstract}

\section{Introduction}
Quantizing gauge theories by using the quantum mechanical path integral (PI) has provided a way of computing amplitudes without losing manifest covariance [1,2,3].  It has proved possible to impose the gauge condition
\begin{equation}
\partial \cdot A^a = 0
\end{equation}
on a vector gauge field $A_\mu^a$ and to relate this to the transversality condition
\begin{equation}
k^\mu \Delta_{\mu\nu}^{ab} (k) = 0
\end{equation}
on the vector vector propagator $\Delta_{\mu\nu}^{ab} (k)$.  This is the ``Landau gauge".  Imposing the gauge condition of eq. (1) results in the introduction of a complex Fermionic scalar ghost field $c^a$ [1-4].

For the propagator $\Delta_{\mu\nu , \lambda\sigma} (k)$ associated with a spin-$2$ gauge field $h_{\mu\nu}$ to be transverse and traceless (``TT'') so that 
\begin{subequations}
\begin{align}
k^\mu \Delta_{\mu\nu ,\lambda\sigma} (k) = 0\\
\intertext{and}
\eta^{\mu\nu} \Delta_{\mu\nu ,\lambda\sigma} (k) = 0
\end{align}
\end{subequations}
no single gauge fixing analogous to that of eq. (1) is possible.  It is necessary to supplement the classical action with a non-quadratic gauge fixing action, as discussed in refs. [5-7]. This results in there being two Fermionic and one Bosonic ghost field.

We will review non-quadratic gauge fixing works, illustrating this technique by considering ``m-n'' gauges in which there are two restrictions
\begin{equation}
n \cdot A^a = m \cdot A^a = 0.
\end{equation}
We will also show how this approach can be extended to incorporate three (or more) gauge fixing conditions.

There is a global gauge symmetry associated with the effective action of Faddeev and Popov; this is the BRST symmetry [8,9]. This symmetry has led to the Batalin-Vilkovisky (BV) approach to covariant quantization of gauge fields [10-13].  We outline the BV quantization procedure and show how it can be applied to the first order Einstein-Hilbert action (IEH) in $1 + 1$ dimensions, thereby giving the BRST symmetry of this model. The way in which non-quadratic gauge fixing can be incorporated into the BV procedure is outlined and is then applied to the massless spin-3/2 action.  This approach allows one to employ a propagator for the spin-3/2 fields that is transverse with respect to both $\partial^\mu$ and $\gamma^\mu$.

\section{Non-Quadratic Gauge Fixing} 
If in the standard PI
\begin{equation}
Z[J^A] = \int D\phi_A \exp i \int dx \left(\mathcal{L}_{c1} (\phi_A) + J^A \phi_A\right)
\end{equation}
the classical action $\mathcal{L}_{c1}$ possesses the gauge invariance
\begin{equation}
\delta\phi_A = R_A^i \theta_i 
\end{equation}
then a degeneracy occurs resulting in PI being ill defined.  This degeneracy can be removed by first inserting a factor of
\begin{equation}
1 = \int D\theta_i \delta \left[ F^{iB} \left( \phi_B + R_\theta^j\theta_j\right) - p^i \right] \det \left( F^{iB} R_B^j\right)
\end{equation}
followed by an insertion of 
\begin{equation}
1 = \int D p^i \exp - \frac{i}{2\alpha}\int dx \left[ p^i N_{ij} p^j \right] \mathrm{det}^{1/2} N_{ij} .
\end{equation}
(The fields $\phi_A$, $\theta_i$ and $p^i$ are all taken to be Bosonic.)  Exponentiating $\det\left(F^{iB} R_B^j\right)$ in eq. (7) results in there being a complex Faddeev-Popov (FP) ghost [1-3] while $\det^{1/2} N_{ij}$ results in a real Nielsen-Kallosh (NK) ghost [14-15].  If one makes the shift $\phi_B \rightarrow
\phi_B - R_B^i\theta_i$ then the PI over $\theta_i$ becomes a (divergent) multiplicative factor in the PI; one then can perform the PI over $p^i$ using the $\delta$-function in eq. (7). 

In refs. [5-7] it is pointed at the eqs. (7,8) can be generalized by using
\begin{align}
1 = \int D\theta_i^1 D\theta_i^2 & \delta \left[ F^{iB} \left(\phi_B + R_B^j \theta^1_j\right) - p^i \right] \det \left( F^{iB} R_B^j\right)\\
& \times \delta 
\left[ G^{iB} \left(\phi_B + R_B^j \theta^2_j\right) - q^i \right] \det \left( G^{iB} R_B^j\right)\nonumber 
\end{align}
and
\begin{equation}
1 = \int Dp^i Dq^i \exp \frac{-i}{\alpha} \int dx \left[ p^i N_{ij} q^j \right] \det N_{ij}. 
\end{equation}
The two determinants in eq. (9) result in a pair of complex FP ghosts, while the determinant in eq. (10) results in a complex NK ghost.  In addition, there is a Bosonic ghost $\theta_i = \theta_i^2 - \theta_i^1$ that arises after the shift of integration variable $\phi_A \rightarrow \phi_A - R_A^i \theta_i^1.$  In ref. [5] it is shown that the TT propagator for the massless spin-2 field is obtained by taking the $\rho \rightarrow 0$ limit of the gauge fixing Lagrangian
\begin{equation}
\mathcal{L}_{gf} = \frac{-1}{\rho} \int dx \left(\partial_\mu h^{\mu\nu}\right) \left(\partial^\lambda h_{\lambda\nu} - \partial_\nu h_\lambda^\lambda\right);
\end{equation}
no quadratic fixing Lagrangian has this property.  Similarly, one could consider the gauge fixing conditions of eq. (4) in Yang-Mills theory.  In this case, since the non-Abelian gauge transformation is 
\begin{equation}
\delta A_\mu^a = \left( \partial_\mu \delta^{ab} + gf^{apb} A_\mu^p \right) \theta^b \equiv D_\mu^{ab} \theta^b 
\end{equation}
eqs. (9,10) lead to an effective action (taking $N_{ij} = \delta_{ij} \partial^2$)
\begin{align}
\mathcal{L}_{eff} = \mathcal{L}_{c1}(A) & - \frac{1}{\alpha}(m \cdot A^a)\partial^2(n \cdot A^a) + \overline{n}^a \partial^2 n^a \nonumber \\
 + \overline{c}^a & (m \cdot D^{ab})c^b + \overline{d}^a(n \cdot D^{ab})d^b \nonumber \\
& + m\cdot A^a n\cdot D^{ab} \theta^b.
\end{align}
In eq. (13) $n^a$, $c^a$ and $d^a$ are complex Fermionic ghost fields while $\theta^a = - \alpha (\theta^{2a} - \theta^{1a})$ is a real Bosonic ghost field.  We note that there is now a mixed propagator for $A_\mu^a$ and $\theta^a$.

We note that eq. (7) implicitly assumes that $F^{iB}$ is such that the gauge function $\theta_j$ can be chosen such that $\phi_B + R_\theta^j \theta_j$ satisfies $F^{iB}(\phi_B + R^j_B\theta_j) - p^i = 0$; similarly in eq. (9) we make the corresponding assumptions on each of the two independent functions $\theta_j^1$ and $\theta_j^2$.  Also, we can see that even though in eq. (9) (and eq. (14) below) we place additional restrictions through use of $\delta$-functions on the gauge fields $\phi_B$, we anticipate that the same number of physical degrees of freedom propagate as if one were to use the conventional expression of eq. (7); this is on account of having introduced a Bosonic ghost field $ \theta_i = \theta_i^2 - \theta_i^1$ as well as a second Fermionic FP ghost.  These new ghosts ensure that even though there are now multiple gauge conditions, the number of propagating degrees of freedom remains unaltered.

Eqs. (9, 10) can be generalized to become
\begin{align}
I = \int D\theta_i^1  D\theta_i^2 D\theta_i^3 & \delta \left[ E^{iB} \left(\phi_B + R_B^j \theta_j^1\right) - p^i\right] \det \left(E^{iB}R_B^j\right)\nonumber \\
& \delta \left[ F^{iB} \left(\phi_B + R_B^j \theta_j^2\right) - q^i\right] \det \left(F^{iB}R_B^j\right)\nonumber \\
& \delta \left[ G^{iB} \left(\phi_B + R_B^j \theta_j^3\right) - r^i\right] \det \left(G^{iB}R_B^j\right)
\end{align}
and
\begin{align}
1 = \int Dp^i Dq^i Dr^i & \exp \frac{-i}{\alpha}\int dx \left[ p^i N^1_{ij} q^j + q^i N_{ij}^2 r^j \right. \\
&\left. + r^i N^3_{ij} p^j\right] \det N_{ij}^1 \det N_{ij}^2 \det N_{ij}^3 .\nonumber
\end{align}
Following the steps outlined above that follow from eqs. (9,10), eqs. (14,15) lead to an effective action that involves three FP ghosts, two real Bosonic ghosts and three NK ghosts.

We now consider how the BV formalism is related to an effective action that involves non-quadratic gauge fixing.

\section{ The BV Formalism}

The BV formalism can be applied directly to the Lagrangian; this avoids the loss of manifest covariance inherent when one considers the Hamiltonian [10-13].  There are numerous reviews of this approach to the quantization of gauge theories among them refs. [12,13,16,17].  We will briefly sketch those aspects of the BV formalism that we need.

If the gauge transformation of eq. (6) has the algebra
\begin{equation}
\frac{\delta R_A^i}{\delta \phi_B} R_B^j - \frac{\delta R_A^j}{\delta \phi_B} R_B^i = R_A^k f_k^{ij}.
\end{equation}
We consider cases where the structure functions $f_k^{ij}$ are independent of $\phi_A$.  Accompanying $\phi_A$ is an ``anti-field'' $\phi^{A\ast}$; ghost fields $c_i$ and $\overline{c}_i$ are associated with the ``anti-ghost fields'' $c^{i\ast}$ and $\overline{c}^{i\ast}$ respectively.  (For a more detailed description of anti-fields, see refs. [10-13, 16, 17].) The field, anti-field action turns out to be 
\begin{equation}
S = S_{c1} (\phi_A) + \phi^{A\ast} R_A^i c_i - \frac{1}{2} c^{i\ast} f_i^{jk} c_jc_k + \overline{c}^{i\ast} N_{ij} A^j
\end{equation}
where $A_i$ is a ``Nakanishi-Lautrup`` (NL) field [18,19].  A properly chosen ghost function $\Psi(\phi_A, c_i, \overline{c}_i)$ can be used to eliminate anti-fields occurring in eq. (17)
\begin{equation}
\phi^{A\ast} = \frac{\delta\Psi}{\delta\phi_A},\quad c^{i\ast} = \frac{\delta\Psi}{\delta c_i}, \quad \overline{c}^{i\ast} = \frac{\delta\Psi}{\delta\overline{c}_i}.
\end{equation}
The BRST transformations that leave $S$ invariant are
\begin{equation}
\delta\phi_A = R_A^i c_i, \quad \delta c_i = - \frac{1}{2} f^{jk}_i c_jc_k .
\end{equation}

To illustrate this, let us consider the IEH action in $1 + 1$ dimensions which can be written as
\begin{equation}
S_{c1} = \frac{1}{2} \int d^2x\; h^{\mu\nu} \left[ G_{\mu\nu ,\lambda}^\lambda + 
G_{\lambda\mu}^\lambda G_{\sigma\nu}^\sigma - G_{\sigma\mu}^\lambda G_{\lambda\nu}^\sigma\right]
\end{equation}
and has the gauge invariance [20,21]
\begin{subequations}
\begin{align}
\delta h^{\mu\nu} = - \left(\epsilon^{\mu\rho} h^{\nu\sigma} + \epsilon^{\nu\rho}h^{\mu\sigma}\right) \theta_{\rho\sigma} \quad \left(\epsilon^{01} = 1 = -\epsilon^{10}\right)\\
\delta G_{\mu\nu}^\lambda = - \epsilon^{\lambda\rho}\theta_{\mu\nu ,\rho} - \epsilon^{\rho\sigma} \left( G_{\mu\rho}^\lambda \theta_{\nu\sigma} + G_{\nu\rho}^\lambda \theta_{\mu\sigma}\right) 
\end{align}
\end{subequations}
as can be determined by using the techniques of refs. [22,23].  (This invariance is not diffeomorphism invariance.)  From eq. (21) we can read off the expressions $R_A^i$ of eq. (6); eq. (16) then gives the structure functions
\begin{align}
f_{\alpha\beta , \gamma\delta}^{\;\;\;\;\;\;\;\;\lambda\sigma} = \frac{-1}{4} \bigg[ \epsilon_{\alpha\gamma} \delta_\beta^\lambda \delta_\delta^\sigma + \epsilon_{\beta\gamma}\delta_\alpha^\lambda\delta_\delta^\sigma + 
\epsilon_{\alpha\delta}\delta_\beta^\lambda\delta_\gamma^\sigma +
\epsilon_{\beta\delta}\delta_\alpha^\lambda \delta_\gamma^\sigma \nonumber \\
  + \epsilon_{\alpha\gamma} \delta_\beta^\sigma \delta_\delta^\lambda + \epsilon_{\beta\gamma}\delta_\alpha^\sigma\delta_\delta^\lambda + 
\epsilon_{\alpha\delta}\delta_\beta^\sigma\delta_\gamma^\lambda +
\epsilon_{\beta\delta}\delta_\alpha^\sigma\delta_\gamma^\lambda\bigg].
\end{align}
The action of eq. (17) becomes
\begin{align}
S = S_{c1} \left(h^{\mu\nu}, G_{\mu\nu}^\lambda\right) + \int d^2x \bigg[ -2h_{\mu\nu}^\ast \left( \epsilon^{\mu\rho} h^{\nu\sigma} c_{\rho\theta}\right) - G_\lambda^{\mu\nu\ast} \big(\epsilon^{\lambda\rho} c_{\mu\nu , \rho}\nonumber \\
+ 2\epsilon^{\rho\sigma} G_{\mu\rho}^\lambda c_{\nu\sigma}\big) + c^{\alpha\beta\ast} \epsilon_{\alpha\gamma} +\overline{c}^{\alpha\beta\ast} A_{\alpha\beta} \bigg] .
\end{align}
A suitable gauge fixing function is
\begin{equation}
\Psi = \int d^2x \;\overline{c}^{\alpha\beta} \left( \epsilon_{\lambda\sigma} G_{\alpha\beta}^{\lambda , \sigma} + \frac{\alpha}{2} A_{\alpha , \beta}\right)
\end{equation}
as the gauge condition
\begin{equation}
\epsilon_{\lambda\sigma} G_{\alpha , \beta}^{\lambda , \sigma} = 0
\end{equation}
has proved useful in showing that no perturbative radiative corrections arise in this model [24].  From eq. (19), we see that the BRST transformations associated with the action of eq. (23) are
\begin{subequations}
\begin{align}
\delta h^{\mu\nu} &= - \left(\epsilon^{\mu\rho} h^{\nu\sigma} + \epsilon^{\nu\rho}h^{\mu\sigma}\right)c_{\rho\sigma} \\
\delta G_{\mu\nu}^\lambda &= - \epsilon^{\lambda\rho}c_{\mu\nu ,\rho} - \epsilon^{\rho\sigma}\left( G_{\mu\rho}^\lambda c_{\nu\sigma} + G_{\nu\rho}^\lambda c_{\mu\sigma}\right) \\
\delta c^{\lambda\sigma} &= \epsilon_{\alpha\beta} c^{\alpha\lambda}c^{\beta\sigma}.
\end{align}
\end{subequations}

We now will relate the BV approach to the quantization of gauge theories to having a non-quadratic gauge fixing term in the effective action, discussed in the preceding section.  This involves the introduction of two ghost fields $c_i$ and $d_i$; eq. (17) is replaced by 
\begin{align}
S = S_{c1} (\phi_A) & + \phi^{A\ast} R_A^i(c_i + d_i) - \frac{1}{4}\left(c_\alpha^\ast + d_\alpha^\ast\right) f_{\beta\gamma}^\alpha \left( c^\beta + d^\beta \right)
\left( c^\gamma + d^\gamma \right)\nonumber \\
&- \frac{2}{\alpha} \left( \overline{c}^{i\ast} N_{ij} B^j + \overline{d}^{i\ast} N_{ij} A^j  \right).
\end{align}
We now take the gauge fixing function to be
\begin{equation}
\Psi = \overline{c}_i \left( F^{iA}\phi_A + A^i\right) + 
\overline{d}_i \left( G^{iA}(\phi_A + R^j_A\theta_j) + B^i\right).
\end{equation}

Eliminating the anti-fields in eq. (27) by using eq. (18) results in
\begin{align}
S = S_{c1}(\phi_A) + & \left( \overline{c}_i F^{iA} +  \overline{d}_i G^{iA} \right) R_A^j (c_j + d_j) \\
& - \frac{2}{\alpha} \bigg[ \left( f^{iA} \phi_A + A^i \right) N_{ij} B^j + A^iN_{ij} \left( G^{jA} \left(\phi_A + R_A^k\theta_k\right) + B^j\right)\bigg]. \nonumber
\end{align}
If one were to now quantize this field anti-field action by performing the PI over $\phi_A$, $c_i$, $\overline{c}_i$, $d_i$, $\overline{d}_i$, $\theta_i$, $A^i$ and $B^i$, one can eliminate $A^i$ and $B^i$ by using their equations of motion
\begin{subequations}
\begin{align}
F^{Ai} \phi_A + 2 A^i &= 0 
\intertext{and}
G^{Ai}\left(\phi_A + R_A^j \theta_j\right) + 2B^i &= 0
\end{align}
\end{subequations}
and make the change of variable
\begin{subequations}
\begin{align}
|\!\!\!C_{\pm i} & = c_i \pm d_i \\
\overline{|\!\!\!C}_\pm^i & = \left(\overline{c}_j F^{Aj} \pm \overline{d}_j G^{Aj}\right)R_A^i
\end{align}
\end{subequations}
in the PI.  Since the Jacobean of the transformation of eq. (28) is
\begin{equation}
\det F^{iA} R_A^j \det G^{iA} R_B^j
\end{equation}
we see that we recover the effective action in the PI that follows from eqs. (9, 10).

\section{Spin-$3/2$}

We have been working with Bosonic fields $\phi_A$.  If $\phi_A$ were to be Fermionic, then the ghost fields $c$, $d$, $\overline{c}$ and $\overline{d}$ would be Bosonic and $\theta$ would be Fermionic.  Taking $\phi_A$ to be Fermionic is appropriate when discussing the massless spin-$3/2$ gauge field $\psi_\mu$ [25].  We now will further demonstrate the utility of non-quadratic gauge fixing by examining the propagator for this field.  The action for this field possesses the gauge invariance 
\begin{equation}
\delta \psi_\mu = \partial_\mu\epsilon
\end{equation}
where $\epsilon$ is a spin-$1/2$ gauge function.  We employ the conventions
\begin{subequations}
\begin{align}
\gamma^{\alpha\beta}& = \frac{1}{2} \left[ \gamma^\alpha, \gamma^\beta \right]\\
\gamma^{\alpha\beta\gamma} &= \frac{1}{2} \left\lbrace \gamma^\alpha, \gamma^{\beta\gamma} \right\rbrace\\
\left\lbrace \gamma^\alpha , \gamma^\beta\right\rbrace = 2\eta^{\alpha\beta}& =
2 \;\mathrm{diag} (+---) 
\end{align}
\end{subequations}
so that
\begin{subequations}
\begin{align}
\gamma_\lambda \gamma^{\lambda\nu\rho}& = 2\gamma^{\nu\rho}= -  \gamma^{\rho\nu\lambda} \gamma_\lambda\\
\gamma^{\lambda\nu\rho} p_\nu & = \frac{1}{2}\left( \gamma^\lambda \gamma \cdot p \gamma^\rho - \gamma^\rho \gamma \cdot p\gamma^\lambda \right)\\
& = \gamma \cdot p \left[ \eta_T^{\lambda\tau} \left( \eta_{\tau\sigma} - \gamma_\tau \gamma_\sigma\right)\eta_T^{\sigma\rho}\right]\\
& \;\;\;\;\;\;\;\;\left(\eta_T^{\lambda\tau} \equiv \eta^{\lambda\tau} - p^\lambda p^\tau/p^2\right)\nonumber \\
\intertext{and}
\left( \gamma^{\kappa\mu\lambda} p_\mu \right) \eta_{\lambda\sigma} \left(p_\nu \gamma^{\sigma\nu\rho}\right) &= /\!\!\!p \left(\gamma^{\kappa\mu\rho} p_\mu \right) - 2p^2 \eta_T^{\kappa\rho}.
\end{align}
\end{subequations}

The kinetic term for a free massless spin-$3/2$ gauge field is proportional to $\overline{\psi}_\lambda \gamma^{\lambda\nu\rho} p_\nu \psi_p$.  If we take the gauge fixing to be quadratic and proportional to $\alpha(\overline{\psi}\cdot\gamma)\gamma \cdot p (\gamma \cdot \psi)$ then the propagator satisfies 
\begin{equation}
S^\kappa_{\,\,\,\,\lambda} \left( - \gamma^{\lambda\nu\rho} p_\nu + \alpha \gamma^\lambda\gamma \cdot p \gamma^\rho \right) = \eta^{\kappa p}
\end{equation}
which leads to 
\begin{equation}
S^{\kappa\lambda} = \frac{1}{p^2}\left[ \frac{1}{2} \gamma^{\kappa\nu\lambda} p_\nu - \left(2 + \frac{1}{\alpha}\right) \frac{p^\kappa \gamma \cdot p p^\lambda}{p^2} \right].
\end{equation}
(If $\alpha = -\frac{1}{2}$ we recover the propagator used in ref. [26].)

For no value of $\alpha$ does $S^{\kappa\lambda}$ satisfy the conditions
\begin{equation}
\gamma_\kappa S^{\kappa\lambda} = 0
\end{equation}
or
\begin{equation}
p_\kappa S^{\kappa\lambda} = 0.
\end{equation}
However, the massive spin-$3/2$ field satisfies the constraints [27]
\begin{equation}
\gamma \cdot \psi = \partial \cdot \psi = 0
\end{equation}
and so the conditions of eqs. (38,39) are the analogue of the TT gauge for a spin-$2$ gauge field as a massive spin-2 field $h_{\mu\nu}$ satisfies
\begin{equation}
\partial_\mu h^{\mu\nu} = \eta_{\mu\nu}h^{\mu\nu} = 0.
\end{equation}
To have the conditions of eqs. (38,39) satisfied we take the gauge fixing action to be proportional to the non-quadratic gauge fixing Lagrangian
\begin{equation}
\xi \left(\overline{\psi} \cdot\gamma \right)\left(p \cdot \psi \right) .
\end{equation}
Using eq. (35) it follows that
\begin{align}
\frac{1}{p^2} \bigg[  & \left(\frac{\xi - 3}{\xi + 6} \right)\gamma^{\kappa\mu\lambda} p_\mu + \frac{2(2\xi + 3)}{(3\xi - 4)(\xi + 6)} p^\kappa\gamma^\lambda\\
& + \frac{3}{\xi + 6} \gamma^\kappa p^\lambda + \frac{2\xi + 3}{\xi + 6}
\gamma \cdot p \eta^{\kappa\lambda} - \frac{2}{\xi(\xi + 6)(3\xi - 4)}\nonumber \\
&\;\;\;\;- \frac{2(3\xi^3-2\xi^2-5\xi + 6)}{\xi(\xi + 6)(3\xi - 4)} \frac{\gamma \cdot p\, p^\kappa p^\lambda}{p^2}\bigg]\nonumber \\
& \;\;\;\;\eta_{\lambda\sigma} \left[ \gamma^{\sigma\nu\rho} p_\nu + \xi\left(p^\sigma \gamma^\rho + p^\rho \gamma^\sigma\right)\right] = \eta^{\kappa\rho}.\nonumber
\end{align}
One can read off from eq. (43) the propagator $S^{\kappa\lambda}$; in the $\xi \rightarrow \infty$ limit it becomes
\begin{equation}
S^{\kappa\lambda} = \frac{1}{p^2}\left(\gamma^{\kappa\nu\rho} p_\nu + 2\eta_T^{\kappa\rho}\right);
\end{equation}
from eq. (35) it follows that eqs. (38,39) are satisfied.

\section{Discussion}

In the preceding sections, we have considered how the standard FP procedure for treating the PI approach to quantizing gauge theories can be extended to incorporate non-quadratic gauge fixing Lagrangians.  The effective action arrived at by this approach is also derived by employing the BV quantization procedure for the case in which the structure function for the gauge algebra of eq. (16) is field independent.  We have demonstrated by BV approach by applying it to the IEH action in $1 + 1$ dimensions.  The free massless spin-$3/2$ action has also been considered; we have shown how using a non-quadratic gauge fixing Lagrangian is needed in this model to give rise to a propagator $S^{\mu\nu}$ that is transverse both in $\partial_\mu$ and the matrix $\gamma_\mu$.

\section*{Acknowledgements}

Roger Macleod had a helpful suggestion.

\end{document}